\documentclass[aps,twocolumn]{revtex4}
\usepackage{amsmath}
\usepackage{bm}
\usepackage{epsfig}
\usepackage{amssymb}

\newcommand{\B}{\mbox{\tiny B}}
\newcommand{\ti}{\tilde}

\newcommand{\nl}{\nonumber \\}
\newcommand{\la}{\langle}
\newcommand{\ra}{\rangle}
\newcommand{\Sec}[1]{Sec.\,\ref{#1}}
\newcommand{\App}[1]{Appendix\;\ref{#1}}

\newcommand{\be}{\begin{equation}}
\newcommand{\ee}{\end{equation}}
\newcommand{\bea}{\begin{eqnarray}}
\newcommand{\eea}{\end{eqnarray}}
\newcommand{\bsube}{\begin{subequations}}
\newcommand{\esube}{\end{subequations}}

\newcommand{\Eq}[1]{Eq.\,(\ref{#1})}
\newcommand{\Eqs}[1]{Eqs.\,(\ref{#1})}
\newcommand{\Fig}[1]{Fig.\,\ref{#1}}

\newcommand{\ind}{{\sf n}}
\newcommand{\bfalp}{\bm\alpha}

\newcommand{\rhonswap}{\rho_{\sf n}^{{ }_{\{\!\leftrightarrow\!\}}}}
\newcommand{\rhonup}{\rho_{\sf n}^{{ }_{\{\!+\!\}}}}
\newcommand{\rhondown}{\rho_{\sf n}^{{ }_{\{\!-\!\}}}}

\begin{document}
%\begin{frontmatter}
\title{Hierarchical theory of quantum dissipation:
Partial fraction decomposition scheme}

\author{Jian Xu and Rui-Xue Xu}
\email{rxxu@ustc.edu.cn}
\address{Hefei National Laboratory for Physical Science at Microscale,
  University of Science and Technology of China, Hefei, Anhui 230026, China}
\author{Meng Luo and YiJing Yan}
\email{yyan@ust.hk}
\address{Department of Chemistry,
  Hong Kong University of Science and Technology, Kowloon,
  Hong Kong SAR, China}
\date{\today}

\begin{abstract}
  We propose a partial fraction decomposition scheme to
the construction of hierarchical equations of motion
theory for bosonic quantum dissipation systems.
The expansion of Bose--Einstein function in this scheme shows
similar properties as it applies for Fermi function. The
performance of the resulting quantum dissipation theory is
exemplified with spin--boson systems. In all cases we have tested
the new theory performs much better, about an order of magnitude
faster, than the best available conventional theory based on
Matsubara spectral decomposition scheme.
\end{abstract}
\iffalse
\fi
%\end{frontmatter}
\maketitle
\section{Introduction}
\label{thintro}

  It is well established now that
path integral influence functional formalism
\cite{Fey63118,Wei08,Kle09} of quantum dissipation theory (QDT) can
be reformulated in terms of
hierarchical equations of motion (HEOM) \cite{Tan89101,Tan914131,%
Tan06082001,Ish053131,Xu05041103,Xu07031107,Jin08234703,Zhe08184112,Zhe09164708}.
 Compared to its dynamics equivalent
path integral influence functional formalism, the HEOM has the
advantage in both numerical efficiency and applications to various
systems. Moreover, the initial system--bath coupling that is not
contained in the original path integral formalism can now be accounted for via
proper initial conditions to HEOM
\cite{Zhe08184112,Zhe09164708}. As an exact formalism, HEOM is
also nonperturbative. It treats the combined effects of system
anharmonicity, system--bath coupling strength, and multiple memory
time scales on the reduced system dynamics.

 The specific HEOM--QDT construction depends however on
the way of decomposing bath correlation function into its memory
components. Different decomposition schemes are mathematically
equivalent, but have different numerical performance. To illustrate
this issue, we consider only the single dissipation mode case,
in which the system--bath coupling Hamiltonian assumes
$H_{\rm SB}=-QF_{\B}$. The system operator $Q$ here is also called the
dissipation mode, through which the generalized Langevin force
$F_{\B}(t)\equiv e^{ih_{\B}t/\hbar}F_{\B}e^{-ih_{\B}t/\hbar}$ acts
on the system. As exact theory is concerned, the bath correlation
function $C(t-\tau)\equiv \la F_{\B}(t)F_{\B}(\tau)\ra$ is related
to the bath spectral density function $J(\omega)$ via
\be\label{FDT}
   C(t)=  \frac{1}{\pi}\int_{-\infty}^{\infty}\!\!d\omega\,
     e^{-i\omega t} \frac{J(\omega)}{1-e^{-\beta\omega}}.
\ee
This is the fluctuation--dissipation theorem of bosonic
canonical ensembles, where $\beta\equiv \hbar/(k_BT)$ denotes the
inverse temperature.

 The common strategy of memory decomposition
used for HEOM construction is based on the Matsubara series
expansion of Bose (also called Bose--Einstein) function in \Eq{FDT}. It
assumes (setting $x=\beta\omega$)
\be \label{laurent}
 \frac{1}{1-e^{-x}} \approx \frac{1}{2}
  +\frac{1}{x} +
   \sum_{m=1}^{N}
   \Big(\frac{1}{x+i2\pi m}+\frac{1}{x-i2\pi m}\Big).
\ee
This is the Matsubara spectral decomposition (MSD) scheme.
It is exact when $N\rightarrow \infty$.
It resolves the memory contents
of bath correlation function with Matsubara frequencies
($2\pi m/\beta$; $m\geq 1$), together with the poles of $J(z)$ in
lower plane (Im\,$z<0$).
In practical use, the residue $\delta C(t)$, due to the deviation of
\Eq{laurent} from the exact one, is treated by white noise
\cite{Shi09084105}.
The resulting HEOM formalism has been applied to electron/population
transfer and optical spectroscopy systems
\cite{Shi09164518,Che09094502,Xu09NJP,Xu09HQME}.

  In this work, we explore
the partial fraction decomposition (PFD) scheme. This scheme was
originally proposed for Fermi function in the study of electronic
dynamics/structure systems
\cite{Gag98399,Goe991085,Goe9317573,%
Nic9414686,Nic9712805,Oza07035123,Cro09073102}. The motivation
behind is that MSD is well--known to be of slow convergence. The PFD of
Bose function will be carried out in a similar manner as that
proposed recently by Croy and Saalmann for the Fermi function
\cite{Cro09073102}.
 We present the PFD results on Bose function in \Sec{thPFD}
and the corresponding HEOM--QDT construction in \Sec{thHEOM}.
Their derivations are given in \App{Append1}
and \App{Append2}, respectively.

  Numerical performance of the new HEOM--QDT will be
exemplified with a model spin--boson Drude dissipation system
in \Sec{thnum}.
Included are also comparisons with the best available
MSD--based conventional HEOM theory. The latter is implemented with
the well--established
Markovian residue correction method \cite{Tan06082001,Ish053131}
that has improved significantly the performance of MSD--based HEOM.
Nevertheless, the PFD--based
HEOM theory without residue correction still performs
much better, about 20 times faster, than the best available MSD--based approach.
Finally we conclude this work.

\section{Partial fraction decomposition scheme}
\label{thPFD}

\subsection{Bose--Einstein function in decomposition}
\label{thpfdA}

The PFD scheme starts with the identity
\be\label{boson}
  \frac{1}{1-e^{-x}}=\frac{1}{2}+\frac{1}{2}\frac{\cosh(x/2)}{\sinh(x/2)}\,,
\ee
followed by Taylor's expansions of the numerator function
$\cosh(x/2)$ to the $(2N)^{\rm th}$--order and the denominator
$\sinh(x/2)$ to the $(2N+1)^{\rm th}$--orders, respectively.
PFD leads to Bose
function the expression (cf.\ \App{Append1})
\be\label{finalexpan}
  \frac{1}{1-e^{-x}}\approx\frac{1}{2}+\frac{1}{x}+
  \sum_{j=1}^N\Big(
    \frac{1}{x+2\sqrt{\xi_j}}+\frac{1}{x-2\sqrt{\xi_j}}
  \Big).
\ee
The involving parameters $\{\xi_j; j=1,\cdots N\}$ are the roots
of the denominator polynomial, $\sum_{n=0}^{N}\xi^n/(2n+1)!=0$.
They can be determined as the eigenvalues
of an $N\times N$ matrix whose elements are
\be\label{Amat_eig}
  A_{mn}=2m(2m+1)\delta_{m+1,n}-2N(2N+1)\delta_{mN}.
\ee
Note that the Fermi function counterpart is similar
\cite{Cro09073102}, just replacing $+1$ with $-1$ in both
parentheses in \Eq{Amat_eig}. The derivation of the above formalism
is detailed in \App{Append1}.

\begin{figure}
\includegraphics[width=0.9\columnwidth]{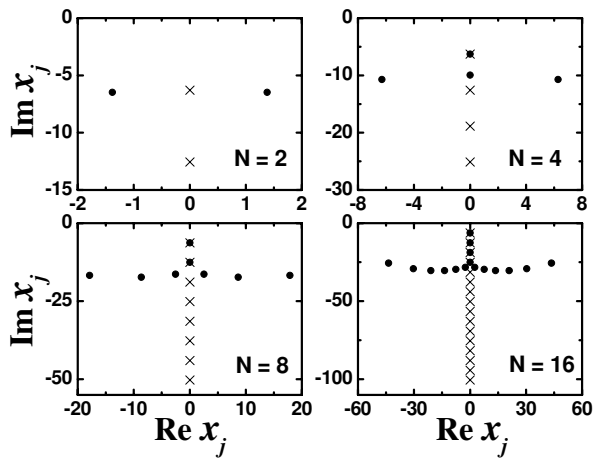}
\caption{Poles of the PFD expansion (solid circle) at the
expansion order $N$=2, 4, 8, 16. Included for comparison are also the
Matsubara poles (cross). Poles are complex--conjugate paired,
and only those in lower plane are shown.
}
\label{fig_pole}
\end{figure}

\begin{figure}
\includegraphics[width=0.9\columnwidth]{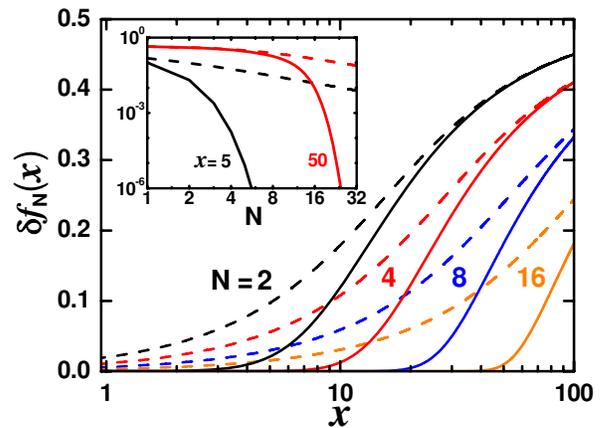}
\caption{The difference $\delta{f}_N(x)\equiv
{f}_{\rm BE}(x)-{f}_{N}(x)$,
between the exact Bose--Einstein function and its
PFD expansion approximation [\Eq{finalexpan}], as the function of $x$ at $N$=2 (black),
4 (red),  8 (blue), and 16 (orange). The inset is for $\delta{f}_N(x)$
as function of $N$, at $x$=5 (black) and 50 (red). The
Matsubara counterparts are also shown by dashed curves.}
\label{fig_diff}
\end{figure}

  Figure \ref{fig_pole} depicts the poles $x_j\equiv \pm 2\sqrt{\xi_j}$
of the PFD scheme at different orders. Shown are only those $N$ poles
in lower plane, and their complex conjugates (in upper plane) are implied.
The distribution pattern of the total $2N$ poles
is similar to that of Fermi function \cite{Cro09073102}.
There are not only pure imaginary poles, which
are mostly close to the Matsubara poles, but also complex poles with
nonzero real parts. This is right the feature for the efficiency of
PFD scheme.

 Figure \ref{fig_diff} depicts the deviation of the approximations
from the exact Bose function.
%%%%
The curves there can be used to estimate the required order of $N$
for the PFD--HEOM dynamics, by considering the effective system and
bath frequencies.
It is clearly seen that the PFD expansion well
overlaps with the exact result, within the range $|x|<4N$
(i.e., $|\omega|<4Nk_BT/\hbar$), as
estimated, although for
the case of Fermi function by Croy and Saalmann recently \cite{Cro09073102}.

\subsection{Correlation function in decomposition}
\label{thpfdB}

 For the later construction of HEOM formalism, we shall
expand the bath correlation function $C(t)$ in an exponential series
\cite{Xu07031107}. This can be achieved via contour integration
method. It recasts \Eq{FDT} via analytic continuation as
\be\label{FDT_contour}
   C(t)=  \frac{1}{\pi}\oint dz\,
     e^{-iz t} \frac{J(z)}{1-e^{-\beta z}},
\ee
and evaluates it by Cauchy's residue theorem.
The contour of integration encloses the lower--half plane
for the required $C(t>0)$. Denote
\be\label{Ctgen}
  C(t) \equiv C_{0}(t)+C_{\rm B}(t),
\ee
for the contributions from the poles of spectral density $J(z)$
and Bose function, respectively. Note that the
antisymmetry property of bosonic spectral
density, $J(-\omega)=-J(\omega)$, reads now
\be\label{antiJz}
 J(-z^{\ast}) = - J^{\ast}(z).
\ee

  Consider first the Bose function contribution $C_{\rm B}(t)$.
The total $N$ PFD poles in lower plane
are either pure imaginary or true complex numbers in pairs,
$\{z_j, -z^{\ast}_j\}$, as depicted in \Fig{fig_pole}. With separation of
pure imaginary and true complex values, we denote these
$N$ poles as
\be\label{zs_def}
  z_s = \left\{\begin{array}{cl}
     -i\gamma_s; \quad               & s=1,\cdots,N_r
   \\ -(\omega_s+i\gamma_s);  \quad  & s=N_r+1,\cdots,N_p
     \end{array}\right.,
\ee
together with $-z^{\ast}_s=\omega_s-i\gamma_s$ in lower plane;
thus, $N_p\equiv(N+N_r)/2$.
The parameters $\gamma_s$ and $\omega_s$ are all positive.
The corresponding residues are $({\rm Res})_{z_s}= e^{-iz_st}J(z_s)$
and $({\rm Res})_{(-z^{\ast}_s)}= e^{iz^{\ast}_st}J(-z^{\ast}_s)
 =-[({\rm Res})_{z_s}]^{\ast}$. The last identity
is inferred from \Eq{antiJz}.
We have therefore
\begin{align}\label{CBt}
   C_{\rm B}(t)
 =  \sum_{s=1}^{N_r} c_s e^{-\gamma_st}
   +\!\!\sum_{s=N_r+1}^{N_p}\!\!\! e^{-\gamma_st}
  [a_s\!\cos(\omega_st)\!-\!b_s\!\sin(\omega_st)].
\end{align}
It decomposes $C_{\rm B}(t)$ into a total of
$N$ components.
The involving coefficients are all real and defined as
\be\label{abc}
   c_s\equiv \frac{2}{i\beta}J(-i\gamma_s), \ \
   a_s+ib_s \equiv \frac{4}{i\beta}J[-(\omega_s+i\gamma_s)].
\ee
The fact that $c_s$ is real can be readily proved via its definition and \Eq{antiJz}.
Thus, $C_{\rm B}(t)$ is always a real function.

  The $C_0(t)$ component of bath correlation function,
which is complex in general, arises
from the poles of spectral density $J(z)$.
In principle it would not
depend on PFD/MSD scheme; but
in practice it will,
for the issue of consistency to be discussed later.
In the following HEOM construction,
we consider the Drude model,
\be \label{aDrudeJ}
    J(\omega) = \frac{\eta\omega_c\omega}{\omega^2+\omega_c^2} ,
\ee
where $\eta$ is the system-bath coupling strength and $\omega_c$
the cut-off frequency.
The Drude spectral density function has only one pole in lower plane.
It results in
\be\label{C0}
  C_0(t) = c_0 e^{-\omega_c t},
\ee
with
$$
  c_0 = -i\eta\omega_c
     \Big(\frac{1}{1-e^{-\beta z}}\Big)_{z=-i\omega_c} .
$$
The above $C_0(t)$ expression is exact. However,
a consistency issue arises as the Bose function
used in evaluating $C_{\rm B}(t)$
via \Eq{CBt} involves the PFD approximation.
The inconsistency here may be problematic in implementation.
For example, the divergence of the above exact
$C_0(t)$ at $\beta\omega_c=2\pi m$
cannot be cancelled out by the approximated
$C_{\rm B}(t)$ of \Eq{CBt}.
 To overcome this problem, we evaluate $c_0$ also
using PFD of \Eq{finalexpan} with the same finite $N$.
\be\label{c0def}
 c_0 = -i\eta\omega_c\Big(\frac{1}{2}+\frac{1}{\beta z}+
  \frac{1}{\beta}\sum_{j=1}^N\frac{2z}{z^2-z^2_j}\Big)_{z=-i\omega_c}.
\ee
The resulting $C_0(t)$ does {\it practically} depend on the
scheme of expansion for Bose function.
 We shall emphasize the importance of
the aforementioned consistent treatment
between $C_0(t)$ and $C_{\rm B}(t)$.
We have recently demonstrated it
with different approximation schemes \cite{Xu09HQME}.
We have also confirmed this issue
with the present PFD scheme,
not just for $C(t)=C_0(t)+C_{\rm B}(t)$, but more
importantly for the numerical performance of
the resulting HEOM dynamics to be presented below.

\section{Hierarchical equations of motion}
\label{thHEOM}

\subsection{HEOM--QDT in PFD scheme}
\label{thheomA}

 The generic form of HEOM--QDT reads \cite{Xu07031107}
\be \label{dotrhon}
 \dot\rho_{\ind} =-(i{\cal L} +{\Gamma_{\ind}} )\rho_{\ind}
  +\rhonswap +\rhondown + \rhonup .
\ee
Here, ${\cal L}\,\cdot\,\equiv \hbar^{-1}[H,\cdot\,]$
is the reduced system Liouvillian;
$\Gamma_{\ind}$ is the decay constant;
$\rho_{\ind}$, $\rhonswap$, $\rhondown$,
and $\rhonup$ are well--defined auxiliary density operators (ADOs)
in the system subspace.
The reduced system density operator  of
primary interest, defined as
$\rho(t)\equiv {\rm Tr}_{\rm bath}\rho_{\rm total}(t)$,
is just $\rho(t)\equiv\rho_{\sf 0}(t)$,
with
 $\rho_{\sf 0}^{{ }_{\{\!\leftrightarrow\!\}}}
 =\rho_{\sf 0}^{{ }_{\{\!-\!\}}}=\Gamma_{\sf 0} =0$.
The labeling index $\ind$ for a generic ADO $\rho_{\ind}$
consists of a set of nonnegative integers, which are arranged
in relation to the individual components
of bath correlation function in a given decomposition scheme.
%%%
Let $\ind=\{n_0,n_1,\cdots,n_{N}\}$ and
$n_0+n_1+\cdots+n_{N}=\tilde n$.
The latter is used to define the tier of $\rho_{\ind}$.
%%%
 The last three
terms in \Eq{dotrhon} describes how a given
$\rho_{\ind}$ of $\ti n^{\rm th}$ tier depends on its associated
ADOs of same tier  ($\rhonswap$) and
neighboring tiers
($\rho_{\ind}^{{}_{\{\!\pm\!\}}}$), respectively.

  The HEOM--QDT formalism is exact and nonperturbative,
assuming only Gaussian bath statistics.
It is equivalent to the Feynman--Vernon
path integral influence functional theory of
reduced system density operator dynamics.
Moreover, it also supports the incorporation of the
initial system--bath correlation
through appropriate initial
$\rho_{\ind\neq{\sf 0}}(t_0)\neq 0$ conditions.
HEOM resolves the combined effect of the coupling bath strength
and memory contents, as they are decomposed,
on the reduced system dynamics.

The specific form of theory depends on
the way of decomposing the bath correlation function $C(t)$.
For Drude dissipation in the PFD scheme, $C(t)$ [\Eq{Ctgen}]
has been decomposed into $(N+1)$ components in
\Eqs{CBt} and (\ref{C0}).
Therefore, the ADO labeling index $\ind$ assumes
\be \label{indexn}
 \ind=\!\left\{n_{s=0,1,\cdots,{N_r}};
    n_{s=N_r+1,\cdots,N_p}; {\bar n}_{s=N_r+1,\cdots,N_p}
 \!\right\}.
\ee
The composite $(N+1)$ nonnegative integers are
the leading orders of individual components of $C(t)$,
involved in the specified $\rho_{\ind}$.
Specifically, $n_{0}$ is for the Drude
component $C_0(t)$ of \Eq{C0},
and the other $N$ integers are for the $N$ components of
$C_{\rm B}(t)$ of \Eq{CBt}, respectively.

 In \App{Append2}, we present the standard approach
to the corresponding HEOM formalism, based on the
present PFD scheme.
The final results are summarized as follows.
 The parameter $\Gamma_{\ind}$ in \Eq{dotrhon}
collects all damping factors. It reads (setting $\gamma_0\equiv\omega_c$):
\be\label{Gamn}
  \Gamma_{\ind} = \sum_{s=0}^{N_r} n_s \gamma_s +
      \sum_{s=N_r+1}^{N_p} (n_s+\bar n_s) \gamma_s\,.
\ee
It arises from the derivatives of individual exponential components of $C(t)$,
involved in $\rho_{\ind}$.

 The swap term  $\rhonswap$ describes how
a given $\rho_{\ind}$ depends on
its associated ADOs of same tier. It reads
\begin{align}\label{swap}
 \rhonswap=&\sum_{s=N_r+1}^{N_p}\!\omega_s\Big[
       ( a_s / b_s )\sqrt{ n_s (\bar n_s+1) | b_s / a_s | }\
       \rho_{{\ind}_s^\rightarrow}
\nl &  \qquad\quad\
    -  ( b_s / a_s )\sqrt{ \bar n_s (n_s+1) | a_s / b_s | }\
       \rho_{{\ind}_s^\leftarrow} \Big],
\end{align}
%%%
It arises from the derivatives of the sine and cosine
components of $C_{\rm B}(t)$ [\Eq{CBt}] involved in $\rho_{\ind}$.
The involving index
${\ind}_s^\rightarrow$ differs from $\ind$ by changing
$(n_s,\bar n_s)$ to $(n_s-1,\bar n_s+1)$ and
${\ind}_s^\leftarrow$ by changing
$(n_s,\bar n_s)$ to $(n_s+1,\bar n_s-1)$.

 The last two terms in \Eq{dotrhon} describes the
tier--down and tier--up contributions.
They  are given by
\begin{align}\label{rhondown}
 \rhondown=&
    -\frac{i}{\hbar} \sum_{s=0}^{N_r}  \sqrt{ n_s / |c_s|}   \Bigl(
    c_s Q \rho_{{\ind}_s^-} - c_s^\ast \rho_{{\ind}_s^-} Q  \Bigr)
\nl &
    -\frac{i}{\hbar} \sum_{s=N_r+1}^{N_p}  a_s \sqrt{ n_s / |a_s|}
    \bigl[Q,\rho_{{\ind}_s^-}\bigr],
\end{align}
\begin{align}\label{rhonup}
 \rhonup=&
    -\frac{i}{\hbar} \sum_{s=0}^{N_r}
     \sqrt{(n_s+1)|c_s|}\,\bigl[Q,\rho_{{\ind}_s^+}\bigr]
\nl &
    -\frac{i}{\hbar} \sum_{s=N_r+1}^{N_p}
    \Big\{\sqrt{(n_s+1)|a_s|}\,  \bigl[Q,\rho_{{\ind}_s^+}\bigr]
\nl &\qquad\qquad
%    -\frac{i}{\hbar}   \sum_{s=N_r+1}^{N_p}
    + \sqrt{(\bar n_s+1)|b_s|}\,    \bigl[Q,\rho_{\bar{\ind}_s^+}\bigr]
    \Big\}.
\end{align}
%%%
They depend explicitly on the system dissipation mode $Q$.
The involving index
${\ind}_s^\pm$ differs from ${\ind}$ by changing the
specified $n_s$ to $n_s \pm 1$,
and $\bar{\ind}_s^\pm$ similarly.

Apparently, the boundary conditions of
 $\rho_{\sf 0}^{{ }_{\{\!\leftrightarrow\!\}}}
 =\rho_{\sf 0}^{{ }_{\{\!-\!\}}}=\Gamma_{\sf 0} =0$
are satisfied.
All coefficients/parameters arising from the Bose part $C_{\rm B}(t)$ of
bath correlation,
$\{a_s,b_s,c_s,\gamma_s,\omega_s\}_{s>0}$,
are real; the last two are positive.
The Drude damping parameter is set to be $\gamma_0=\omega_c$.
Only $c_0$ [\Eq{c0def}] is complex.

\subsection{Remarks on implementation}
\label{thheomB}

 To facilitate locating a specified ADO,
we also like to have a working index scheme
to map a set of $(N+1)$ ordered multiple indices,
${\ind}\equiv \{n_0,n_1,\cdots,n_N\}$,
to an integer $j_{\ind}$, such that
$\rho_{\ind}\equiv \rho_{j_{\ind}}$.
%%%
That $\rho_{\ind}$ is an ${\tilde n}^{\rm th}$--tier ADO
if its $n_0 + \cdots + n_N\equiv \ti n$.
The number of ADOs at a given tier is
$\frac{(N+\tilde n)!}{N!\,\ti n!}$,
while the total number of ADOs up to level $L$ is
\be\label{calN}
 {\cal N} = \sum_{\ti n=0}^{L} \frac{(N+\tilde n)!}{N!\,\ti n!}
    \equiv
    \Big\{\begin{array}{c} L \\ N \end{array} \Big\}.
\ee
The second identity serves also the definition of
$\left\{^m_{\,n}\right\}$
for abbreviated notion,
with the boundary values of $\left\{^{m<0}_{\ \ n}\right\}=0$ and
$\left\{^{0}_{n}\right\}=\left\{^{m}_{\;0}\right\}=1$.
The working index $j_{\ind}\equiv j_{n_0\cdots n_N}=0,\cdots,{\cal N}-1$
can then be
\be\label{appindex}
  j_{\ind}
=   \Big\{\begin{array}{c} \ti n-1 \\ N \end{array} \Big\} +
    \sum_{q=0}^N \Big\{\begin{array}{c}
      n_{q+1}+\cdots+n_{N}-1 \\ N-q
     \end{array} \Big\}.
\ee
It sorts the index ${\ind}$ by tiers and followed by subindices,
so that $j_{{\ind}=\{0,\cdots,0\}}=0$,
$j_{{\ind}=\{1,0,\cdots,0\}}=1$,
$\cdots$, and
$j_{{\ind}=\{0,\cdots,0,L\}}={\cal N}-1$.

 In practice, HEOM should be truncated properly
at finite level of hierarchy $L$ and
decomposition order $N$.
By far the truncation of $N$
goes with convergency, but
that of $L$ are carried out
effectively and automatically.
Apparently, the issue of truncation
is closely related to
central processing unit (CPU) time and memory cost of computation.
The number ${\cal N}$ of total ADOs goes with a combinatory law,
like the complete configuration interaction treatment
in quantum mechanics. Shi et al.\ have recently proposed an efficient,
accuracy--controlled, dynamics filtering algorithm \cite{Shi09084105,Shi09164518}.
The basic idea behind is the observation
that only a very small fraction of total ADOs
play roles in HEOM propagation.
The filtering algorithm sets a specific $\rho_{\ind}(t_j)=0$ if
its matrix elements amplitudes are all
smaller than the pre-chosen error tolerance.
To validate this simple algorithm, all ADOs
including the reduced density matrix of the primary interest
should have a uniform filtering--algorithm error tolerance.
The present HEOM formalism [\Eqs{dotrhon}--(\ref{rhonup})]
has been scaled properly to meet this requirement.
In connection with the stochastic field approach to
Gaussian--Markovian dissipation, the scaled ADOs
are just the expansion coefficients, with the normalized
harmonic wave functions used as the basis set for resolving
the diffusive bath field \cite{Shi09164518}.
The filtering algorithm keeps only those necessary ADOs
according to the selected error tolerance.
Apparently, it also automatically truncates the
level of hierarchy {\it on--the--fly} during numerical
propagation.

\section{Numerical performance and concluding remarks}
\label{thnum}

\begin{figure}
\includegraphics[width=0.9\columnwidth]{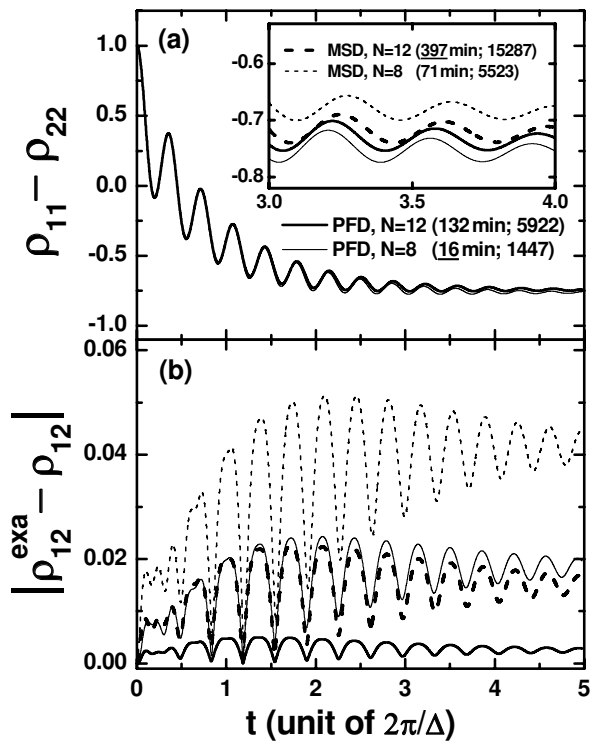}
\caption{Time evolution of a spin--boson system density matrix:
(a) the population difference $\rho_{11}(t)-\rho_{22}(t)$;
(b) the deviation of off-diagonal $\rho_{12}(t)$ from the exact.
The initial conditions are all $\rho_{\ind}(0)=0$;
except for $\rho_{\ind=\sf 0}(0)=\rho(0)$ where $\rho_{11}(0)=1$
and $\rho_{22}(0)=\rho_{12}(0)=0$.
The specified values of (CPU\,time;\,${\cal N}_{\rm eff}$)
highlight the implementation cost.
Especially the two underlined values indicate
that PFD--HEOM converges about 25 times faster than
MSD--HEOM. See text for details.
}
\label{fig_dyn}
\end{figure}

  The performance of the present PFD--based HEOM dynamics is
exemplified with \Fig{fig_dyn}
for a spin--boson system $H=\epsilon\sigma_z+\Delta\sigma_x$,
with the dissipation mode of $Q=\sigma_z$.
The parameters for the system and coupling bath spectral density
are $(\epsilon, \eta, \omega_c)=(1, 0.5, 5)$ in unit of $\Delta$,
with $\Delta=1660$\,cm$^{-1}$ and $T=298$\,K.
The result of PFD--HEOM with $N=12$
in \Fig{fig_dyn} is practically converged,
within 0.005 from that with $N=16$.
The latter is treated as
the exact in \Fig{fig_dyn}(b).
The maximum tier of survival ADOs
is found to be $L_{\rm max}=8$, with the
filtering algorithm error tolerance of
$10^{-7}$;
see  \Sec{thheomB} or Ref.\,\cite{Shi09084105}.

Included for comparison are also the results of the best
available MSD--based conventional HEOM theory.
This reference theory is augmented with the well--established
Markovian residue correction (MRC) \cite{Tan06082001,Ish053131},
in which $\delta C(t)\equiv C_{\rm exact}(t)-C_{\rm MSD}(t)$
is approximated by white noise for its
effect on the MSD--based HEOM dynamics.
Without additional implementation cost,
this treatment significantly reduces
the required number $N$ for converged MSD
dynamics \cite{Tan06082001,Ish053131,Shi09084105,Shi09164518,Che09094502,Xu09NJP,Xu09HQME}.
This remarkable MSD--based feature remains true
for the system in \Fig{fig_dyn}.
However, MRC does not work well with the present
PFD--HEOM theory.
%A proper residue correction
%if needed for the new theory
%is yet to be established.
Therefore the comparison shown in \Fig{fig_dyn} is between
PFD--HEOM without residue correction
versus MSD--HEOM with MRC,
the best available reference as we know.

 Performance is reported in terms of
(CPU\,time;\,${\cal N}_{\rm eff}$).
The CPU time is by a single Intel(R) Xeon(R) processor\,@3.00\,GHz,
with the fourth-order Runge--Kutta propagator and time--step of 0.0015\,fs.
Another parameter ${\cal N}_{\rm eff}$ records
the largest number of ADOs ever survived
during propagation with the filtering algorithm.
%%%
Recall that the ADOs in PFD--HEOM is of the
maximum survival tier level of $L_{\rm max}=8$.
In contrast, a converged MSD--HEOM dynamics requires
the level of about 20 tiers.

  For the purpose of comparison, however,
we set $L=9$ in all calculations, including both PFD and MSD schemes.
 Apparently, ${\cal N}_{\rm eff}$ in either scheme
is orders of magnitude smaller than the total number of ADOs,
which is ${\cal N}=497420$ for $N=12$
or ${\cal N}=48620$ for $N=8$, respectively.
The PFD--HEOM is of smaller ${\cal N}_{\rm eff}$,
about $25\sim40\%$ of its MSD counterpart
with same $L$ and $N$.
  The enhanced filtering efficiency in PFD scheme here
is closely related to its complex poles, rather than
only pure imaginary ones (\Fig{fig_pole}).
The complex poles lead to oscillatory decomposition components
of bath correlation function [\Eq{CBt}],
and result in oscillatory cancelation
in PFD--HEOM dynamics.
This right property of the PFD scheme
may account for the relatively small number
of survival ADOs after filtering algorithm.

 Performance of PFD--HEOM dynamics should be based on the CPU
time versus its MSD counterpart of the same quality.
As mentioned earlier, however, the converged MSD dynamics
for the present system of study
is too expensive to be worth here.
We rather choose for assessment by
a pair of similar quality but approximated results: PFD($N=8$) versus MSD($N=12$),
with the CPU time of 16 minutes versus 397 minutes,
respectively, see \Fig{fig_dyn}.
We have carried out a series of dynamics simulations with different system and bath parameters.
All results show that the performance of PFD--HEOM
is superior over its MSD counterpart by at least an order
of magnitude.

 In summary, we have constructed the PFD scheme to Bose function
and HEOM--QDT. The superiority of PFD over MSD is apparent. The
complex PFD poles lead to not just the Bose function expansion more
efficient and accurate, but also the HEOM construction more compact.
The resulting HEOM dynamics converges
with smaller $L$ and $N$,
and also accommodates better with the propagation filtering algorithm.
These factors contribute the performance of PFD--based HEOM theory,
which converges a magnitude (about 20 times) faster than its MSD
counterpart.

\section*{Acknowledgement}
 Support from the National Natural Science Foundation of China
 (20533060 and 20773114),
 National Basic Research Program of China (2006CB922004),
 and RGC Hong Kong (604508 and 604709) is acknowledged.

\appendix
\section{Derivation of Bose function in PFD expansion}
\label{Append1}

  The derivation of \Eqs{finalexpan} and (\ref{Amat_eig}) starts with
the Taylor's expansions
$\cosh(y)\approx\sum_{n=0}^N y^{2n}/(2n)!$
and $\sinh(y)\approx  \sum_{n=0}^N y^{2n+1}/(2n+1)!$
for the numerator and denominator of the last term
in \Eq{boson}, respectively.
We have then
\be\label{boson_PFD0}
  \frac{1}{1-e^{-x}} \approx \frac{1}{2}+\frac{1}{x} + B_N(x/2)\,.
\ee
where
\begin{align}\label{Bxdef}
  B_N(y)
&=\dfrac{1}{2}
  \frac{\sum_{n=0}^N y^{2n}/(2n)!}
       {\sum_{n=0}^N y^{2n+1}/(2n+1)!} - \frac{1}{2y}
\nl&=
  \frac{\sum_{n=1}^N \frac{n}{(2n+1)!}\, y^{2n-1}
  }{
    \sum_{n=0}^N \frac{1}{(2n+1)!}\, y^{2n}
  }
\nl &\equiv
 \frac{1}{2}\sum_{j=1}^N \Big(\frac{b^+_j}{y+\sqrt{\xi_j}}
   +\frac{b^-_j}{y-\sqrt{\xi_j}} \Big),
\end{align}
with $\{y_j=\pm \sqrt{\xi_j}; j=1,\cdots,N\}$ being
the roots of the denominator polynomial; i.e.,
\be\label{denom_poly}
  \sum_{n=0}^N \frac{1}{(2n+1)!}\,\xi^n = 0.
\ee
The PFD coefficient $b^{\pm}_j$ can be determined via
\begin{align}\label{PFD_coef}
 b^{\pm}_j &=
   2 \lim_{y\pm\sqrt{\xi_j}=0}\left[(y\pm\sqrt{\xi_j})B_N(y)\right]
\nl&=
  2 \lim_{y=0}\left[yB_N(y\mp\sqrt{\xi_j})\right]
\nl&=\left.
 \frac{\left[2y\sum_{n=1}^N \frac{n}{(2n+1)!} (y\mp\sqrt{\xi_j})^{2n-1}\right]'
  }{
    \left[\sum_{n=0}^N \frac{1}{(2n+1)!}(y\mp\sqrt{\xi_j})^{2n}\right]'
  }\right\vert_{y=0}
\nl&=
  \frac{\sum_{n=1}^N \frac{2n}{(2n+1)!} (\mp\sqrt{\xi_j})^{2n-1}
      }{\sum_{n=0}^N \frac{2n}{(2n+1)!}(\mp\sqrt{\xi_j})^{2n-1}}
\nl &= 1.
\end{align}
The PFD expression of Bose function,
\Eq{finalexpan}, is obtained by substituting \Eqs{Bxdef} and (\ref{PFD_coef})
into \Eq{boson_PFD0}, with $x=2y$.

 To convert the roots of polynomial to eigenvalue problem,
as \Eq{Amat_eig}, let us consider the general case of
\be\label{polyN}
  a_0 + a_1 \xi + \cdots + a_N \xi^{N} = 0.
\ee
We search for the proper matrix
${\bm A}\equiv\{A_{mn}\}$, with
the eigenvector the
form  of ${\bf v}=[c_1,c_2\xi,\cdots,c_N\xi^{N-1}]^T$
that converts the eigenequation ${\bm A}{\bf v}=\xi{\bf v}$ to
\be\label{eigA2}
  \sum_{n=1}^{N} A_{mn}c_n\xi^{n-1}=c_m\xi^{m};
 \ \ \ m=1,\cdots,N.
\ee
By comparing \Eqs{polyN} and (\ref{eigA2}) and setting
$c_N=a_N$, we have the following nonzero elements:
$A_{Nn} = -a_{n-1}/c_n$ and
$A_{m,m+1}=c_m/c_{m+1}$ for $m\neq N$;
others are all zeroes. Therefore,
\be\label{Amat}
  A_{mn} = \frac{c_m}{c_{m+1}}\delta_{m+1,n}
    - \frac{a_{n-1}}{c_n} \delta_{mN} ,
\ee
with $c_n$ being arbitrary nonzero parameters, except for $c_N=a_N$.
%%%

  In particular, we choose $c_{n}  = a_{n-1} a_{N}/a_{N-1}$,
if every $a_j\neq 0$.
The boundary condition of $c_N=a_N$ is also satisfied.
We can recast \Eq{Amat} as
\be\label{Amat_final}
  A_{mn} = \frac{a_{m-1}}{a_{m}}\delta_{m+1,n}
    - \frac{a_{N-1}}{a_N} \delta_{mN} .
\ee
It recovers \Eq{Amat_eig} for the polynomial  \Eq{denom_poly},
where $a_n=1/(2n+1)!$.

\section{Derivation of HEOM--QDT in PFD scheme}
\label{Append2}

 The following derivation of the HEOM in \Sec{thheomA}
is carried out via the well--established
Calculus--On--Path--Integral Influence--Generating--Functional
(COPI--IGF) algebra \cite{Xu07031107}.
It starts with the path integral influence
functional for quantum Gaussian dissipation \cite{Fey63118,Wei08,Kle09},
followed by consecutive time derivatives
to resolve  in a hierarchical manner the
involving memory contents \cite{Tan89101,Tan914131,Xu05041103}.
Unlike the HEOM theory that can be expressed in operator level,
the path integral formalism has to go with representation.
 Let $\{|\alpha\ra\}$ be a generic basis set in the system subspace
and ${\bm \alpha}\equiv(\alpha,\alpha')$ for abbreviation,
such that
$\rho({\bm \alpha},t) \equiv \rho(\alpha,\alpha'\!,t)
  \equiv \la\alpha|\rho(t)|\alpha'\ra$.
Introduce the reduced Liouville--space propagator ${\cal U}$ via
\be \label{arhotPI_def}
 \rho({\bm \alpha},t)
 \equiv \!\int\!d{\bm \alpha}_0\, {\cal U}({\bm \alpha},t;{\bm \alpha}_0,t_0)
 \rho({\bm \alpha}_0,t_0).
\ee
Its path--integral expression reads
\be \label{acalGPI}
   {\cal U}(\bfalp,t;\bfalp_0,t_0)
 = \int_{\bfalp_0[t_0]}^{\bfalp[t]}   \!\!  {\cal D}{\bfalp} \,
     e^{ iS[\alpha]/\hbar} {\cal F}[\bfalp]
     e^{-iS[\alpha']/\hbar} .
\ee
$S$ and ${\cal F}$ are the action and influence functionals, respectively.
For Gaussian bath interactions,
the latter assumes the Feynman--Vernon form
that can be recast as \cite{Fey63118,Xu07031107}
\be \label{FV_FPhiA}
   {\cal F}[\bfalp] =
      \exp\Bigl\{-\frac{i}{\hbar}\int_{t_0}^t\!\!d\tau\,{\cal A}[\bfalp(\tau)]\,
   {\cal B}(\tau;\{\bfalp\})\Bigr\},
\ee
where
\be\label{calA}
  {\cal A}[\bfalp(t)] = Q[\alpha(t)]-Q[\alpha'(t)],
\ee
\bsube\label{calBall}
\be\label{calB}
  {\cal B}(t;\{\bfalp\}) =
 -\frac{i}{\hbar}\big[B(t;\{\alpha\})- B'(t;\{\alpha'\})\big],
\ee
with
\be
\begin{split}
 B(t;\{\alpha\}) &\equiv \int_{t_0}^{t}\!d\tau\,
    C(t-\tau) Q[\alpha(\tau)],
\\
 B'(t;\{\alpha'\}) &\equiv \int_{t_0}^{t}\!d\tau\,
 C^{\ast}(t-\tau) Q[\alpha'(\tau)].
\end{split}
\ee
\esube
Note that ${\cal A}[\bfalp(t)]$ depends only on the local time
of path and assumes readily in operator level as commutator,
${\cal A}\,\cdot = [Q, \,\cdot\,]$.
  In contrast, the functional ${\cal B}(t;\{\alpha\})$ contains
memory and does not have a simple operator--level correspondence.
The COPI--IGF algebra provides a proper hierarchy
to resolve the memory contents involved in ${\cal B}(t;\{\alpha\})$.

 To proceed, we decompose ${\cal B}(t;\{\alpha\})$
according to the decomposition of bath correlation function $C(t)$
[\Eq{Ctgen} with \Eqs{CBt} and (\ref{C0})]. We have
\be\label{calB_decom}
  {\cal B}=\sum_{s=0}^{N_r}{\cal B}_{s}
    +\sum_{s=N_r+1}^{N_p} ({\cal B}_{s}+\bar{\cal B}_{s}),
\ee
where (noting that $c_0$ is complex while others are real)
\bsube\label{allBs}
\begin{alignat}{2}
{\cal B}_s &\equiv -\frac{i}{\hbar}(c_sB_s-c_s^\ast B'_s);
   &\quad s&=0,1,\cdots,N_r  \, ,
\\
{\cal B}_s  &\equiv -\frac{i}{\hbar} a_s(B_s-B'_s);
  &\quad s&=N_r+1,\cdots,N_p  \, ,
\\
\bar{\cal B}_s&\equiv \frac{i}{\hbar} b_s(\bar B_s-\bar B'_s);
  &\quad s&=N_r+1,\cdots,N_p  \, .
\end{alignat}
\esube
with (denoting $\omega_{s\leq N_r}\equiv 0$)
\be\label{Bs_def}
\begin{split}
  B_s& = \int_{t_0}^{t}\!d\tau\,
     e^{- \gamma_s(t-\tau)}
     \cos[\omega_s(t-\tau)]
     Q[\alpha(\tau)],
\\
     \bar B_s& = \int_{t_0}^{t}\!d\tau\,
     e^{- \gamma_s(t-\tau)}
     \sin[\omega_s(t-\tau)]
     Q[\alpha(\tau)].
\end{split}
\ee
The time derivative of \Eq{Bs_def} reads
\be
\begin{split}
 \partial_t B_s
& = -\gamma_sB_s-\omega_s\bar B_s + Q[\alpha(t)],
\\
  \partial_t \bar B_s
& = -\gamma_s\bar B_s + \omega_s B_s.
\end{split}
\ee
Thus the memory functionals
of \Eq{allBs} satisfy
\bsube \label{dotcalBs}
\be\label{dotcalBsa}
  \partial_t{\cal B}_s=-\gamma_s{\cal B}_s
-\frac{i}{\hbar}\bigl\{c_s Q[\alpha(t)]-c_s^\ast Q[\alpha'(t)]\bigr\},
\ee
for $s=0,1,\cdots,N_r$; while
\begin{align}
  \partial_t {\cal B}_s=&-\gamma_s{\cal B}_s
    +\frac{a_s}{b_s}\omega_s\bar{\cal B}_s
      -\frac{i}{\hbar}a_s {\cal A}[{\bm\alpha}(t)],
\label{dotcalBsb} \\
 \partial_t\bar{\cal B}_s=&-\gamma_s\bar{\cal B}_s
-\frac{b_s}{a_s}\omega_s{\cal B}_s,
\label{dotcalBsc}
\end{align}
\esube
for $s=N_{r}+1,\cdots,N_p$.

 The above equations of motion for
$\{{\cal B}_{s},\bar{\cal B}_s\}$ are closed, with the inhomogeneous terms
depending only on the local time.
This is right the property for
these $\{{\cal B}_{s},\bar{\cal B}_s\}$
defined in \Eq{allBs}
to be the IGFs for the desired hierarchy construction.
The auxiliary influence functionals ${\cal F}_{\ind}$,
with the labeling index ${\ind}$ of \Eq{indexn},
are now obtained via the IGFs as \cite{Xu07031107}
\be\label{acalFsfn}
 {\cal F}_{\ind}=\frac{1}{\sqrt{\sigma_{\ind}}}
 \left[\prod_{s=0}^{N_r}{\cal B}_s^{n_s}\ \cdot
 \!\!  \prod_{s=N_r+1}^{N_p}  \!\!
 ( {\cal B}_s^{n_s} \bar{\cal B}_s^{\bar n_s} )
 \right] {\cal F},
\ee
which specifies the ADO as
$\rho_{\ind}(t)\equiv{\cal U}_{\ind}(t,t_0)\rho(t_0)$ with
\be\label{acalGPIaux}
 {\cal U}_{\ind}(\bfalp,t;\bfalp_0,t_0)
     \equiv\!\int_{\bfalp_0[t_0]}^{\bfalp[t]}
   \!\!\!  {\cal D}{\bfalp} \,
     e^{iS[\alpha]/\hbar} {\cal F}_{\ind}[\bfalp] e^{-iS[\alpha']/\hbar}.
\ee

 Included in \Eq{acalFsfn} is also a proper scaling parameter $\sigma_{\ind}$.
This is for the purpose of applying simple and efficient filtering algorithm
\cite{Shi09084105}
in HEOM propagation.
This parameter scales the specified $\rho_{\ind}$ to be of
not just the same unit, but also about the
same error tolerance as the reduced system
density operator $\rho\equiv\rho_{\sf 0}$ of primary interest.
It is given by \cite{Shi09084105,Shi09164518}:
\be\label{scale}
 \sigma_{\ind}=\prod_{s=0}^{N_r}(|c_s|^{n_s}n_s!)\ \cdot
 \!\! \prod_{s=N_r+1}^{N_p} \!\!
 (|a_s|^{n_s}n_s!\,|b_s|^{\bar n_s}\bar n_s!).
\ee

 The HEOM [\Eqs{dotrhon}--(\ref{rhonup})] can now be obtained readily
by taking the time derivative on $\rho_{\sf n}(t)$
or its propagator ${\cal U}_{\ind}$ of \Eq{acalGPIaux}.
Various terms in \Eqs{dotrhon}
and (\ref{Gamn})--(\ref{rhonup})
result from the derivatives on various components in
\Eqs{acalFsfn} and \Eq{acalGPIaux}, as detailed as follows.
The coherent Liouvillian dynamics part in \Eq{dotrhon}
arises from the derivative of the action functionals
in \Eq{acalGPIaux};
The tier--up $\rhonup$ term [\Eq{rhonup}] is from
 $\partial_t {\cal F}=-\frac{i}{\hbar}{\cal A}({\cal BF})$,
 following by using \Eqs{calB_decom} and (\ref{acalFsfn}),
and the operator--level form of  ${\cal A}\,\cdot = [Q,\,\cdot\,]$;
Finally,
${\Gamma}_{\ind}$, $\rhonswap$,
 and $\rhondown$ [\Eqs{Gamn}--(\ref{rhondown})] collect
the decay, swap, and inhomogeneous terms of \Eqs{dotcalBs}, respectively.
Note that the scaling parameters, $\sigma_{\ind}$ of \Eq{scale},
are also involved.

%\bibliographystyle{elsarticle-num}
%\bibliography{bibrefs}

\begin{thebibliography}{10}
\expandafter\ifx\csname url\endcsname\relax
  \def\url#1{\texttt{#1}}\fi
\expandafter\ifx\csname urlprefix\endcsname\relax\def\urlprefix{URL }\fi
\expandafter\ifx\csname href\endcsname\relax
  \def\href#1#2{#2} \def\path#1{#1}\fi

\bibitem{Fey63118}
R.~P. Feynman, F.~L. \mbox{Vernon, Jr.},
 Ann. Phys. 24 (1963) 118.

\bibitem{Wei08}
U.~Weiss, Quantum Dissipative Systems, World Scientific, Singapore, 2008, 3rd
  ed.

\bibitem{Kle09}
H.~Kleinert, Path Integrals in Quantum Mechanics, Statistics, Polymer Physics,
  and Financial Markets, World Scientific, Singapore, 2009, 5th ed.

\bibitem{Tan89101}
Y.~Tanimura, R.~Kubo,
 J. Phys. Soc. Jpn. 58 (1989) 101.

\bibitem{Tan914131}
Y.~Tanimura, P.~G. Wolynes,
 Phys. Rev. A 43 (1991) 4131.

\bibitem{Tan06082001}
Y.~Tanimura,
 J. Phys. Soc. Jpn. 75 (2006) 082001.

\bibitem{Ish053131}
A.~Ishizaki, Y.~Tanimura,
 J. Phys. Soc. Jpn. 74 (2005) 3131.

\bibitem{Xu05041103}
R.~X. Xu, P.~Cui, X.~Q. Li, Y.~Mo, Y.~J. Yan,
 J. Chem. Phys. 122 (2005) 041103.

\bibitem{Xu07031107}
R.~X. Xu, Y.~J. Yan,
 Phys. Rev. E 75 (2007) 031107.

\bibitem{Jin08234703}
J.~S. Jin, X.~Zheng, Y.~J. Yan,
 J. Chem. Phys. 128 (2008) 234703.

\bibitem{Zhe08184112}
X.~Zheng, J.~S. Jin, Y.~J. Yan,
 J. Chem. Phys. 129 (2008) 184112.

\bibitem{Zhe09164708}
X.~Zheng, J.~S. Jin, S.~Welack, M.~Luo, Y.~J. Yan,
 J. Chem. Phys. 130 (2009) 164708.

\bibitem{Shi09084105}
Q.~Shi, L.~P. Chen, G.~J. Nan, R.~X. Xu, Y.~J. Yan,
 J. Chem. Phys. 130 (2009) 084105.

\bibitem{Shi09164518}
Q.~Shi, L.~P. Chen, G.~J. Nan, R.~X. Xu, Y.~J. Yan,
 J. Chem. Phys. 130 (2009) 164518.

\bibitem{Che09094502}
L.~P. Chen, R.~H. Zheng, Q.~Shi, Y.~J. Yan,
 J. Chem. Phys. 131 (2009) 094502.

\bibitem{Xu09NJP}
J.~Xu, R.~X. Xu, Y.~J. Yan,
 New J. Phys. In press.

\bibitem{Xu09HQME}
R.~X. Xu, B.~L. Tian, J.~Xu, Q.~Shi, Y.~J. Yan, J. Chem. Phys. Submitted.

\bibitem{Gag98399}
F.~Gagel,
 J. Comput. Phys. 139 (1998) 399.

\bibitem{Goe991085}
S.~Goedecker,
 Rev. Mod. Phys. 71 (1999) 1085.

\bibitem{Goe9317573}
S.~Goedecker,
 Phys. Rev. B 48 (1993) 17573.

\bibitem{Nic9414686}
D.~M.~C. Nicholson, G.~M. Stocks, Y.~Wang, W.~A. Shelton, Z.~Szotek, W.~M.
  Temmerman,
   Phys. Rev. B
  50 (1994) 14686.

\bibitem{Nic9712805}
D.~M.~C. Nicholson, X.-G. Zhang,
 Phys. Rev. B 56 (1997) 12805.

\bibitem{Oza07035123}
T.~Ozaki,
 Phys. Rev. B 75 (2007) 035123.

\bibitem{Cro09073102}
A.~Croy, U.~Saalmann,
 Phys. Rev. B 80 (2009) 073102.

\end{thebibliography}

\end{document}